\documentclass[aps,twocolumn,showpacs]{revtex4}
\usepackage{epsfig}
\usepackage{amsbsy}
\usepackage{amsmath}
\usepackage{amsfonts}
\usepackage{graphicx}

\begin{document}

\newcommand{\bqa}{\begin{eqnarray}}
\newcommand{\eqa}{\end{eqnarray}}
\newcommand{\nn}{\nonumber}
\newcommand{\nl}[1]{\nn \\ && {#1}\,}
\newcommand{\erf}[1]{Eq.\ (\ref{#1})}
\newcommand{\dg}{^\dagger}
\newcommand{\ip}[1]{\left\langle{#1}\right\rangle}
\newcommand{\bra}[1]{\langle{#1}|}
\newcommand{\ket}[1]{|{#1}\rangle}
\newcommand{\braket}[2]{\langle{#1}|{#2}\rangle}
\newcommand{\st}[1]{\langle {#1} \rangle}
\newcommand{\etal}{{\it et al.}}
\newcommand{\half}{{\small \frac 12}}

\title{Numerical Analysis of the Capacities for Two-Qubit Unitary Operations}
\author{Dominic W.\ Berry$^1$ and Barry C.\ Sanders$^{1,2}$}
\address{$^1$ Australian Centre for Quantum Computer Technology, Department of
Physics, Macquarie University, Sydney, New South Wales 2109, Australia \\
$^2$ Quantum Information Science Group, Department of Physics and
Astronomy, University of Calgary, Alberta T2N 1N4, Canada}
\date{\today}

\begin{abstract}
We present numerical results on the capacities of two-qubit unitary operations for
creating entanglement and increasing the Holevo information of an ensemble. In all
cases tested, the maximum values calculated for the capacities based on the Holevo
information are close to the capacities based on the entanglement. This indicates
that the capacities based on the Holevo information, which are very difficult to
calculate, may be estimated from the capacities based upon the entanglement, which
are relatively straightforward to calculate.
\end{abstract}
\maketitle

\section{Introduction}
A nonlocal operation is one which operates on distinct subsystems, and can not be expressed
as a tensor product of operations on the individual subsystems. Such operations arise via
an interaction Hamiltonian between the subsystems. Nonlocal operations may be used to
create entanglement between subsystems, and also to perform classical communication. In
fact, it is not possible to achieve either of these tasks without an interaction between
the subsystems. In characterizing quantum operations, it is thus important to determine the
capacities for creating entanglement or performing communication. These capacities provide
a characterization of the strength of the operations \cite{nielsen}.

Shared classical information may be considered to be the classical equivalent of
entanglement. Therefore it is reasonable to consider the process of classical
communication to be the classical equivalent of entanglement creation. One may therefore
expect that there is a close relationship between the capacities for these two tasks. In
fact, it has been shown \cite{bhls,berry1,berry2} that there are a number of
inequalities between the various capacities for classical communication and entanglement
creation. In this paper we make a direct numerical comparison between these capacities.

The capacities of operations for creating entanglement have been studied extensively
\cite{cirac,zanardi,durvid,kradur,kraus,leifer,childs,kraus2}.
It is relatively straightforward to determine the entanglement capability for
infinitesimal operations \cite{durvid}. For finite operations, most results are restricted
to numerical results for certain classes of two-qubit operations \cite{leifer}. We study
the same classes of operations here, and the results we present for the entanglement
reproduce those given in Ref.\ \cite{leifer}, except for some data points where we believe
our results are more accurate.

The capacities for classical communication were initially considered for simple
operations, such as the CNOT and SWAP operations \cite{eisert,collins}. It is more difficult to
consider the classical communication for general operations, because general operations
can not be used to perform perfect communication. Ref.\ \cite{bhls} introduced asymptotic
capacities, where the average communication when the operation is performed a very large
number of times is considered. When the operation is performed a large number of times,
it is possible to use error correcting techniques to reduce the probability of error to
be arbitrarily small.

For the purposes of numerical investigation, it is not practical to use asymptotic
capacities, as the number of variables that would need to be optimized over to obtain
a reasonable approximation is extremely large. A far more practical type of capacity is
that based on the Holevo information of ensembles. For these capacities, only the states
and probabilities in the initial ensemble need be optimized over in the numerical
analysis. In addition, there are connections between the capacities defined via the
change in Holevo information and the asymptotic capacities \cite{bhls}.

Although it is feasible to calculate capacities based on the Holevo information, it is
far more computationally difficult than evaluating the entanglement capacities.
It would therefore be useful if it were possible to use the entanglement capacities
to estimate the capacities based on the Holevo information. In this paper we numerically
study the relationship between these capacities, and show that although they are not equal,
the values we have calculated are quite close.

In Sec.\ \ref{sec:defs} we explain the definitions of the various capacities.
In Sec.\ \ref{sec:inzer} we give numerical results for the communication capacity based
on the Holevo information obtained when the initial ensemble has zero Holevo
information, and compare these capacities to the entanglement that may be created from
initial states that have zero entanglement. Then in Sec.\ \ref{sec:ingen} we give
results for the increase in Holevo information for general initial ensembles, and
compare these capacities to the increase in entanglement for arbitrary initially
entangled states. 
We conclude in Sec.\ \ref{concl}, and in the Appendix give a detailed explanation of
the numerical techniques used to calculate the results presented.

\section{Definitions}
\label{sec:defs}
First we summarize the definitions of the various capacities. Throughout this paper
we divide the system into two subsystems, $A$ and $B$, and denote the Hilbert spaces
by ${\cal H}_A$ and ${\cal H}_B$. The party in possession of subsystem $A$ will be
referred to as Alice and the party in possession of subsystem $B$ will be referred
to as Bob. The subsystems $A$ and $B$ are divided into further subsystems:
\begin{equation}
\label{ancilla}
{\cal H}_A = {\cal H}_{A_U} \otimes {\cal H}_{A_{\rm anc}}, ~~~~~
{\cal H}_B = {\cal H}_{B_U} \otimes {\cal H}_{B_{\rm anc}}.
\end{equation}
The operation $U$ acts only upon ${\cal H}_{A_U} \otimes {\cal H}_{B_U}$, and the
Hilbert spaces ${\cal H}_{A_{\rm anc}}$ and ${\cal H}_{B_{\rm anc}}$ are ancillas.
The ancillas may have dimension that is arbitrarily large but finite.

There are two main ways of defining capacities for entanglement. The first is the
entanglement that may be obtained when the initial state is unentangled
\footnote{Subsystems A and B are not entangled, but there may be entanglement within
the respective subsystems.}
\begin{equation}
\label{enzer}
E_U \equiv \sup_{\ket{\phi}_A\in{\cal H}_A,\ket{\chi}_B\in{\cal H}_B}
E(U\ket{\phi}_A\ket{\chi}_B).
\end{equation}
The quantity $E(\cdots)$ is the entropy of entanglement
$E(\ket{\Psi}) = S[{\rm Tr}_A(\ket{\Psi}\bra{\Psi})]$, where
$S(\rho)=-{\rm Tr}(\rho \log\rho)$. Throughout we employ
logarithms to base 2, so the entanglement is expressed in units of ebits.
The second definition is the maximum increase in entanglement when the initial
state may be an arbitrary pure entangled state.
\begin{equation}
\label{endif}
\Delta E_U \equiv \sup_{\ket{\Psi}\in{\cal H}_A\otimes{\cal H}_B}
\left[ E(U\ket{\Psi}) - E(\ket{\Psi})\right].
\end{equation}
It is also possible to define the capacity in terms of the entanglement of
formation and allow mixed states; we do not separately consider this case, because
allowing mixed states does not alter the capacity \cite{bhls}.

Another alternative definition of the entanglement capacity is based on the
average entanglement that may be obtained in the limit that the operation is
performed an extremely large number of times and the initial state is unentangled
\cite{bhls}. It is shown in Ref.\ \cite{bhls} that this capacity is equal to the
maximum increase in entanglement as defined in Eq.\ (\ref{endif}). Thus the
numerical results presented also apply to the asymptotic entanglement capacity.

In the numerical search, it is not possible to consider ancilla spaces with
arbitrarily large dimension. It is known \cite{nielsen} that the ancilla spaces
need have dimension no larger than the Hilbert spaces ${\cal H}_{A_U}$ and
${\cal H}_{B_U}$ for the capacity with initially unentangled states. It has also
been found numerically \cite{leifer} that the same is true for the capacity
where initially entangled states are allowed.

In the results presented below, we use a fixed dimension for the ancilla spaces.
For simplicity we use equal dimensions on the two ancilla spaces 
${\cal H}_{A_{\rm anc}}$ and ${\cal H}_{B_{\rm anc}}$. We use a superscript on
the capacity when it is necessary to indicate the dimension of the ancilla space
used. For example, $\Delta E_U^{(4)}$ is the maximum change in entanglement when
the ancilla spaces are each of dimension 4. When we refer to multiple results with
different ancilla dimensions we will use a superscript $(*)$.
We will omit the superscript in the case of $E_U$ when the dimensions of the
ancilla spaces are at least as large as those of ${\cal H}_{A_U}$ and
${\cal H}_{B_U}$, because this is known to be sufficient to obtain the capacity
for arbitrarily large ancilla.

The classical communication capacities that we will consider are based upon the
Holevo information of ensembles. An ensemble is a set of states
$\{\ket{\Phi_i}_{AB}\}$ that are supplied with probabilities $p_i$. Each state
$\ket{\Phi_i}_{AB}$ is a pure state shared between Alice and Bob, and Alice
chooses the index $i$. The ensemble is denoted by
${\cal E}=\{ p_i, \ket{\Phi_i}_{AB} \}$. We also define the ensemble of reduced
density matrices possessed by Bob as
\begin{equation}
{\sf E}={\rm Tr}_A{\cal E}=\{ p_i, \rho_i \},
\end{equation}
where $\rho_i = {\rm Tr}_A \ket{\Phi_i}_{AB}\bra{\Phi_i}$.
The Holevo information of the ensemble ${\sf E}$ is given by
\begin{equation}
\chi({\sf E}) = S\left(\sum_i p_i \rho_i\right) - \sum_i p_i S(\rho_i).
\end{equation}
From the Holevo-Schumacher-Westmoreland theorem \cite{holevo,shuwes},
the Holevo information gives the average communication that may be performed
from Alice to Bob by coding over multiple states.

Similarly to the case for entanglement, we may define capacities based on the
maximum change in Holevo information. One definition that we will use is the
maximum final Holevo information when the initial ensemble has zero Holevo
information. For the initial ensemble, we have an initial state
$\ket{\psi}_{AB}$, and Alice encodes $i$ by applying a local operation $V^{(i)}$.
For the capacity, the supremum is taken over the initial state
$\ket{\psi}_{AB}$, the encoding operations $V^{(i)}$ and the probabilities $p_i$:
\begin{equation}
\chi_U = \sup_{ p_i, V^{(i)}, \ket{\psi}_{AB} }
\chi\left( p_i, {\rm Tr}_A U V^{(i)} \ket{\psi}_{AB} \right),
\end{equation}
where ${\rm Tr}_A \ket{\phi} \equiv {\rm Tr}_A \ket{\phi}\bra{\phi}$. Note that the
notation we are using here differs from Ref.\ \cite{bhls}, where the symbol
$\Delta\chi_U^{(1,\emptyset)}$ was used for this capacity.

We may also define the maximum change in Holevo information when the initial
ensemble is arbitrary:
\begin{equation}
\Delta \chi_U = \sup_{\cal E} \chi ( {\rm Tr}_A U{\cal E} )
- \chi ( {\rm Tr}_A {\cal E} ).
\end{equation}
Here we are using the notation conventions 
\begin{align}
U{\cal E} &\equiv \{ p_i, U\ket{\Phi_i} \}, \\
{\rm Tr}_X{\cal E} &\equiv \{ p_i, {\rm Tr}_X(\ket{\Phi_i}) \}.
\end{align}
This capacity is equivalent to the capacity $\Delta \chi_U^{(1,*)}$ defined in
Ref.\ \cite{bhls}. As shown in Ref.\ \cite{bhls}, this capacity is equal to the
average entanglement-assisted communication that may be performed from Alice to
Bob. Therefore this quantity may be interpreted as the asymptotic communication
capacity, just as $\Delta E_U$ may be interpreted as the asymptotic
entanglement capability.

One may also interpret $\chi_U$ in terms of asymptotic capacities. The capacity
$\chi_U$ gives the Holevo information after a single application of the operation
$U$. This communication can not actually be performed for a single ensemble; it
is necessary to code over multiple states to perform this average communication.
Therefore $\chi_U$ may be interpreted as the asymptotic communication capacity
if the communication protocol is limited to the relatively simple scheme where
coding is performed over multiple final states. This is equivalent to restricting
all the applications of $U$ to be performed at the same time (on input states
that are not entangled with each other), rather than allowing the output of one
application of $U$ to be used as part of the input to another application of $U$,
as in the general case.

As in the case of the entanglement, it is not possible to use ancilla spaces of
arbitrarily large dimension in the numerical search. In addition, it is not
possible to use arbitrarily large numbers of states in the ensemble. In the
results presented below, we perform calculations for restricted ensembles
where there is a fixed number of states in the ensemble and a fixed dimension
for the ancilla spaces (the ancilla spaces are again taken to be of equal
dimension). We use superscripts on the capacities to indicate the number of
states in the ensemble and the dimension of the ancilla spaces. For example,
$\Delta \chi_U^{(2,4)}$ is the maximum change in Holevo information for two
states in the ensemble and ancillas each of dimension 4. We use a superscript
$(*)$ to refer to multiple capacities with different ancilla dimensions or
ensemble sizes. It must be emphasized that our use of superscripts in this paper
differs from that in Ref.\ \cite{bhls}.

\section{Capacities for zero initial Holevo information}
\label{sec:inzer}
It is clear that the capacity $\chi_U$ is an analogous quantity for
communication to $E_U$ for entanglement. Similarly $\Delta\chi_U$ is analogous
to $\Delta E_U$ for entanglement. In this section we perform a direct numerical
comparison between the two capacities $\chi_U$ and $E_U$. In the next section we
compare the capacities $\Delta\chi_U$ and $\Delta E_U$.

In this paper we concentrate on two-qubit operations. It is not possible to
perform calculations for the entire range of two-qubit operations. To make the
problem feasible, we only consider a limited number of examples of two-qubit
operations. In particular, we consider operations of the form:
\begin{align}
\label{U1}
U_1(\alpha) &=U_d(\alpha,0,0), \\
\label{U2}
U_2(\alpha) &=U_d(\alpha,\alpha,0), \\
\label{U3}
U_3(\alpha) &=U_d(\alpha,\alpha,\alpha),
\end{align}
where
\begin{equation}
\label{simpleU}
U_d(\alpha_1,\alpha_2,\alpha_3) = {\rm e}^{ -i ( \alpha_1
\sigma_1 \otimes \sigma_1 + \alpha_2 \sigma_2 \otimes \sigma_2 + \alpha_3
\sigma_3 \otimes \sigma_3 ) }.
\end{equation}
Here $\sigma_i$ are the Pauli sigma operators.
The operations $U_1$, $U_2$ and $U_3$ correspond to the CNOT, DCNOT and SWAP
families of operations considered in Ref.\ \cite{leifer}.

In order to consider the complete range of two-qubit operations in the case of
entanglement, it would only be necessary to consider operations of the form
(\ref{simpleU}), with $\pi/4\ge\alpha_1\ge\alpha_2\ge\alpha_3\ge 0$ \cite{kraus}.
This derivation relies on the fact that any two-qubit operation may be simplified
to one of the form (\ref{simpleU}) with $\pi/4\ge\alpha_1\ge \pm\alpha_2\ge\alpha_3
\ge 0$ using local operations \cite{durvid,kraus,makhlin,hammer}. In addition to
using local operations, the derivation in Ref.\ \cite{kraus} relies on the fact that
the entanglement capabilities of $U$ and $U^*$ are identical (which implies that all
the $\alpha_i$ may be taken to be positive).

Similarly, for the Holevo information, $\chi({\rm Tr}_A{\cal E})=\chi
({\rm Tr}_A{\cal E}^*)$ and $\chi({\rm Tr}_A U{\cal E})=\chi({\rm Tr}_A
U^*{\cal E}^*)$. Thus the capacities of $U$ and $U^*$ to increase the Holevo
information are identical. Therefore, in order to obtain results for the complete
range of two-qubit operations, it would only be necessary to consider operations of
the form (\ref{simpleU}) with $\pi/4\ge\alpha_1\ge\alpha_2\ge\alpha_3\ge 0$ for both
entanglement and Holevo information. This restriction on the
values of the $\alpha_i$ defines a three dimensional region of values. In this
paper we do not consider the entire region; however, the operations $U_1$, $U_2$
and $U_3$ which we consider form three lines on the boundaries of this region.

\begin{figure}
\centering
\includegraphics[width=0.45\textwidth]{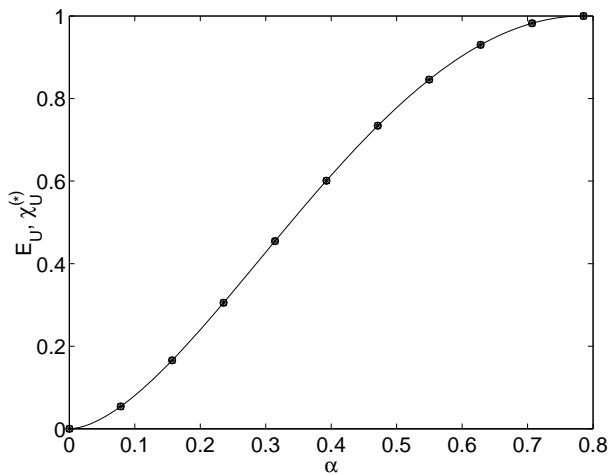}
\caption{The increase in entanglement for no initial entanglement, $E_{U_1}$, and the
increase in Holevo information for zero initial Holevo information, $\chi_{U_1}^{(*)}$.
The values of $E_{U_1}$ are shown as the solid line, and the values of
$\chi_{U_1}^{(2,1)}$, $\chi_{U_1}^{(2,2)}$, $\chi_{U_1}^{(4,2)}$ and
$\chi_{U_1}^{(4,4)}$ are shown as the circles, crosses, plusses and squares,
respectively. These symbols overlap and are not visible separately.}
\label{figU1a}
\end{figure}

Sophisticated numerical maximization techniques were used to find the values of $E_U$ and
$\chi_U^{(*)}$; the details are given in the Appendix.
The numerical results for $E_{U_1}$ and $\chi_{U_1}^{(*)}$ are shown in
Fig.\ \ref{figU1a}. The values of $E_{U_1}$ were determined in intervals of $\pi/400$ for
$\alpha$, and are shown by the solid line. An ancilla dimension of 2 is sufficient to
obtain the asymptotic capacity $E_U$ for all two-qubit operations \cite{nielsen}. In all
the calculations presented for $E_U$ in this section, an ancilla dimension of 2 was
used; we have therefore omitted the superscript on this capacity. In addition, it was
found that, for $U_1$, the same capacity was obtained without ancilla, in agreement with
Ref.\ \cite{kraus}.

The values of $\chi_{U_1}^{(*)}$ were determined in intervals of $\pi/40$ for $\alpha$,
with ancilla dimensions of 1, 2 and 4, and ensembles with 2 or 4 members. An ancilla
dimension of 1 is equivalent to no ancilla; we will take the case of no ancilla to be
an ancilla dimension of 1 for convenience in the notation.
Just as in the case of the entanglement, these results indicate that the maximum final
Holevo information is obtained without ancillas. In addition, only two states in the
ensemble are required. We find no improvement in using up to four states in the
ensemble, and ancillas of dimension up to four. It was found that the maximum increase
in the Holevo information was obtained with equal probabilities when there were two
states in the ensemble. When four states in the ensemble were used, in most cases two
of the probabilities approached zero whereas the other two approached $1/2$.

In addition, note that there is no difference between the results obtained for the
maximum final entanglement and the maximum final Holevo information. It was found
that the two capacities agreed to precision better than one part in $10^{14}$.
Our results do not prove that $E_{U_1}=\chi_{U_1}$, because it is not possible to test
ensembles with arbitrarily large numbers of states or allow the ancilla dimension to
be arbitrarily large. Nevertheless, our results strongly indicate that $E_{U_1}$ is
equal to $\chi_{U_1}$.

\begin{figure}
\centering
\includegraphics[width=0.45\textwidth]{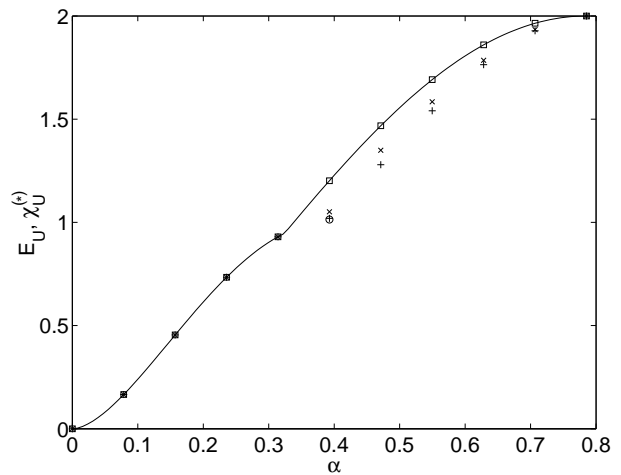}
\caption{The increase in entanglement for no initial entanglement, $E_{U_2}$, and the
increase in Holevo information for zero initial Holevo information, $\chi_{U_2}^{(*)}$.
The values of $E_{U_2}$ are shown as
the solid line, and the values of $\chi_{U_2}^{(4,2)}$, $\chi_{U_2}^{(4,3)}$ and
$\chi_{U_2}^{(4,4)}$ are shown as the plusses, crosses and squares, respectively.
All results for $\chi_{U_2}^{(4,5)}$ are equal to $\chi_{U_2}^{(4,4)}$, and are not shown
be separate symbols.
The data point indicated by the circle is for $\chi_{U_2}^{(4,2)}$ when the
probabilities are restricted to be equal.}
\label{figU2a}
\end{figure}

Numerical results for $E_{U_2}$ and $\chi_{U_2}^{(*)}$ are shown in
Fig.\ \ref{figU2a}. As for $E_{U_1}$, the value of $E_{U_2}$ was determined in intervals of
$\pi/400$ for $\alpha$, and an ancilla dimension of 2 was used in order to obtain the
asymptotic entanglement capability $E_{U_2}$. The values of $\chi_{U_2}^{(*)}$ were determined in
intervals of $\pi/40$ for $\alpha$, and with ancilla dimensions of 2, 3, 4 and 5. In
each case the ensemble consists of four states.

In contrast to the case for the entanglement, there are improvements in using ancilla
dimensions higher than 2 when considering the Holevo information. For an ancilla
dimension of 2, the value of $\chi_{U_2}^{(4,2)}$ is equal to $E_{U_2}$ for the lower values
of $\alpha$, but is less than $E_{U_2}$ for the upper values of $\alpha$. For an ancilla
dimension of 3, the values of $\chi_{U_2}^{(4,3)}$ are larger, but still less than $E_{U_2}$
towards the right side of the plot. For an ancilla dimension of 4, $\chi_{U_2}^{(4,4)}$
is equal to $E_{U_2}$. When the ancilla dimension is increased to 5, the capacity
$\chi_{U_2}^{(4,5)}$ is no larger than $\chi_{U_2}^{(4,4)}$. These results indicate that
an ancilla dimension of 4 may be sufficient to achieve the capacity $\chi_{U_2}$, and that
this capacity is equal to $E_{U_2}$.

Similar calculations were performed with equal probabilities for each of the
states in the ensemble. In most cases the same results were obtained as when
arbitrary probabilities were allowed. However, for one data point for
$\chi_{U_2}^{(4,2)}$ (indicated by the circle in Fig.\ \ref{figU2a}), a larger Holevo
information was obtained when unequal probabilities were allowed. The numerically derived
optimal ensemble had three members with non-negligible probabilities: two probabilities
were approximately $41.39\%$, and one was $17.22\%$.

\begin{figure}
\centering
\includegraphics[width=0.45\textwidth]{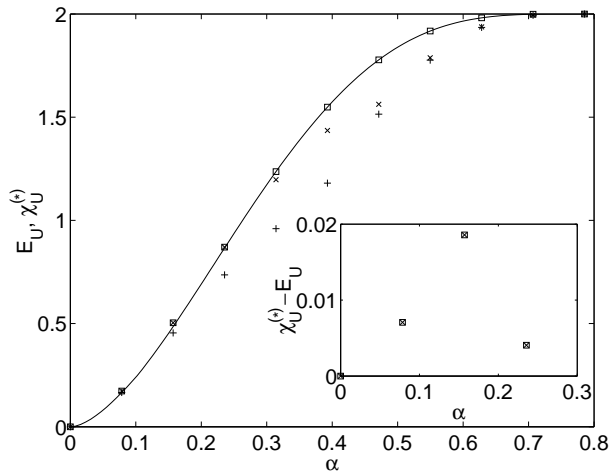}
\caption{The increase in entanglement for no initial entanglement, $E_{U_3}$, and the
increase in Holevo information for zero initial Holevo information, $\chi_{U_3}^{(*)}$.
The values of $E_{U_3}$ are shown as the solid line, and the values of
$\chi_{U_3}^{(4,2)}$, $\chi_{U_3}^{(4,3)}$ and $\chi_{U_3}^{(4,4)}$ are shown as the
plusses, crosses and squares, respectively. All values of $\chi_{U_3}^{(4,5)}$ are
identical to $\chi_{U_3}^{(4,4)}$, and are not shown independently.
The subplot shows the difference $\chi_{U_3}^{(*)}-E_{U_3}$ for $\chi_{U_3}^{(4,3)}$
(crosses) and $\chi_{U_3}^{(4,4)}$ (squares).}
\label{figU3a}
\end{figure}

Results for $E_{U_3}$ and $\chi_{U_3}^{(*)}$ are shown in
Fig.\ \ref{figU3a}. Again the value of $E_{U_3}$ was determined in intervals of $\pi/400$
for $\alpha$, and with an ancilla dimension of 2. The values of $\chi_{U_3}$ were determined
for ancilla dimensions of 2, 3, 4 and 5, and with four states in the ensemble.

As in the case of $\chi_{U_2}^{(4,2)}$, the values of $\chi_{U_3}^{(4,2)}$ are significantly below
$E_{U_3}$ for most of the plot. When the ancilla dimension is increased to 3, the
values of $\chi_{U_3}^{(4,3)}$ are noticeably larger, and closer to $E_{U_3}$. In addition,
three of the values, at $\alpha=\pi/40$, $2\pi/40$ and $3\pi/40$, are actually
larger than $E_{U_3}$ (see the subplot in Fig.\ \ref{figU3a}).
The difference is small, less than $0.02$, but it
is sufficient to demonstrate that $E_{U_3}$ is not equal to $\chi_{U_3}$. For
an ancilla dimension of 4, $\chi_{U_3}^{(4,4)}$ is not smaller than $E_{U_3}$ for any of
the data points. In addition, $\chi_{U_3}^{(4,4)}$ is larger than $E_{U_3}$ for the same
three values of $\alpha$ as for an ancilla dimension of 3. In fact, $\chi_{U_3}^{(4,4)}$
is equal to $\chi_{U_3}^{(4,3)}$ for these three data points.
When the ancilla dimension is increased to 5, $\chi_{U_3}^{(4,5)}$ is no larger than
$\chi_{U_3}^{(4,4)}$, again indicating that an ancilla dimension of 4 is sufficient to achieve
the capacity $\chi_{U_3}$.

To summarize, we have found that, for the operation $U_1$, ancillas do not increase
the capacity $\chi_U$, just as in the case for the entanglement. For the cases $U_2$ and
$U_3$, the capacity $\chi_U$ increases with ancilla dimension up to an ancilla dimension
of 4, but is unchanged when the ancilla dimension is further increased to 5. This is in
contrast to the case for the entanglement, where the maximum entanglement capacity $E_U$
is obtained for an ancilla dimension of 2 \cite{nielsen}.

The results indicate that $\chi_U$ is equal to $E_U$ for the operations $U_1$ and
$U_2$, though it is possible that $\chi_U^{(*)}$ is increased for larger ancilla
dimensions or ensemble sizes. For the operation $U_3$, it is possible to obtain slightly
higher values of $\chi_U^{(*)}$, demonstrating that $\chi_U$ is not equal to $E_U$ for
this operation. In all cases tested, we find that $\chi_U\ge E_U$. In only a small number
of cases have we found that $\chi_U\ne E_U$, and in these cases the differences found are
only small, suggesting that $E_U$ is an excellent, and efficient, estimator of $\chi_U$.

\section{Capacities for arbitrary initial ensembles}
\label{sec:ingen}
Next we consider the capacities $\Delta\chi_U$ and $\Delta E_U$. These capacities
are more general, in that arbitrary initial states or ensembles are allowed.
Analytic results for the relation between these capacities have been derived in
Ref.\ \cite{berry2}. In this reference it is shown that, for two-qubit operations,
$\Delta\chi_U\ge\Delta E_U$. If $\Delta E_U$ may be achieved with a particular
ancilla dimension, then an increase in Holevo information equal to $\Delta E_U$
may be achieved with the same ancilla dimension, and with four members in the
ensemble.

In principle it is possible that there is no ancilla dimension that achieves
$\Delta E_U$, and instead $\Delta E_U$ is approached in the limit of large ancilla
dimension. However, in practice it has been found that, for two-qubit operations,
it appears to be possible to achieve $\Delta E_U$ with an ancilla dimension of 2
\cite{leifer}. This means that it should be possible to achieve an increase in Holevo
information equal to $\Delta E_U$ with an ancilla dimension of 2.

\begin{figure}
\centering
\includegraphics[width=0.45\textwidth]{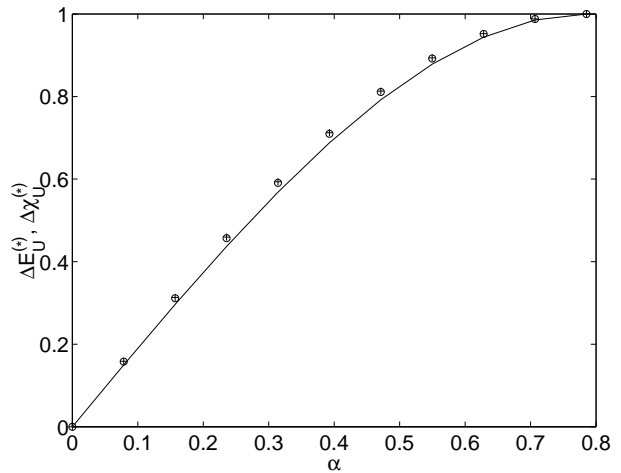}
\caption{The increase in entanglement for arbitrary initial state, $\Delta E_{U_1}^{(*)}$,
and the increase in Holevo information for arbitrary initial ensemble,
$\Delta\chi_{U_1}^{(*)}$. The values of $\Delta E_{U_1}^{(1)}$
and $\Delta E_{U_1}^{(2)}$ are shown as the solid line, and the values of
$\Delta\chi_{U_1}^{(2,1)}$ and $\Delta\chi_{U_1}^{(2,2)}$ are shown as the circles and
plusses, respectively.}
\label{figU1b}
\end{figure}

The capacities $\Delta\chi_{U_1}^{(*)}$ and $\Delta E_{U_1}^{(*)}$ are shown
in Fig.\ \ref{figU1b}. Each capacity was determined in steps of $\pi/40$ in $\alpha$.
It was found that the entanglement capacity $\Delta E_{U_1}^{(*)}$ did not increase beyond
that for no ancilla as the ancilla dimension was increased up to 5, in agreement
with the result given in Ref.\ \cite{leifer}. Throughout this section we use the
superscript asterisk when referring to results for $\Delta E_U^{(*)}$, because it has
not been proven that the asymptotic capacity is achieved for an ancilla dimension of 2.

The capacities $\Delta\chi_{U_1}^{(*)}$ without ancillas and with ancillas of dimension 2 are
shown in Fig.\ \ref{figU1b}. In both cases these capacities are for ensembles with
two states. It is found that, even without ancilla, the capacity
$\Delta\chi_{U_1}^{(*)}$ is greater than the values calculated for $\Delta E_{U_1}^{(*)}$. The only
cases where there is equality are the trivial cases where $\alpha=0$ or $\pi/4$.
These results indicate that there are operations for which there is the strict
inequality $\Delta\chi_U>\Delta E_U$.

\begin{figure}
\centering
\includegraphics[width=0.45\textwidth]{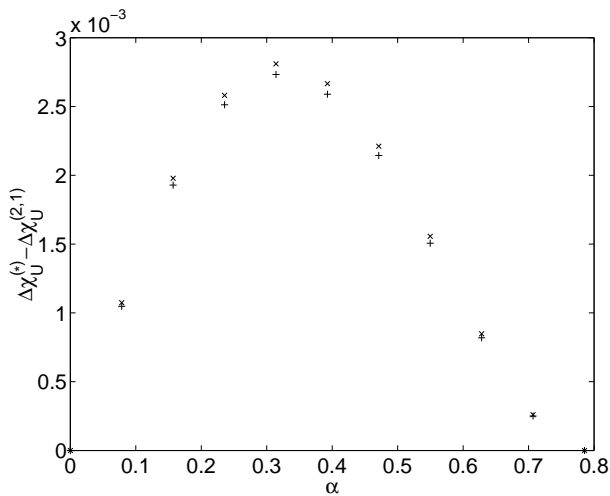}
\caption{The increase in Holevo information for arbitrary initial ensemble,
$\Delta\chi_{U_1}^{(*)}$, relative to that for the case where the ensemble has two members
and there is no ancilla, $\Delta\chi_{U_1}^{(2,1)}$. The values of
$\Delta\chi_{U_1}^{(2,2)}-\Delta\chi_{U_1}^{(2,1)}$ are shown as the plusses and the values
of $\Delta\chi_{U_1}^{(2,3)}-\Delta\chi_{U_1}^{(2,1)}$ are shown as the crosses.}
\label{figU1c}
\end{figure}

In addition, the capacity $\Delta\chi_{U_1}^{(*)}$ is slightly increased by adding an ancilla.
This is not so easily seen in Fig.\ \ref{figU1b}; to make this difference visible, the
differences between the capacities $\Delta\chi_{U_1}^{(*)}$ with ancilla and the capacity with
no ancilla $\Delta\chi_{U_1}^{(2,1)}$ are plotted in Fig.\ \ref{figU1c}. It can be seen
that there is a small but significant increase in $\Delta\chi_{U_1}^{(*)}$ when an ancilla is
allowed.

\begin{figure}
\centering
\includegraphics[width=0.45\textwidth]{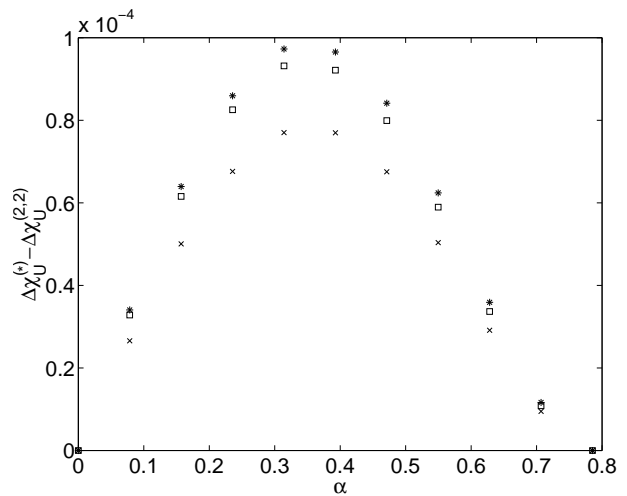}
\caption{The increase in Holevo information for arbitrary initial ensemble,
$\Delta\chi_{U_1}^{(*)}$, relative to that for the case where the ensemble has two members
and the ancilla is of dimension 2, $\Delta\chi_{U_1}^{(2,2)}$. The values of
$\Delta\chi_{U_1}^{(2,3)}-\Delta\chi_{U_1}^{(2,2)}$, $\Delta\chi_{U_1}^{(2,4)}-\Delta\chi_{U_1}^{(2,2)}$
and $\Delta\chi_{U_1}^{(2,5)}-\Delta\chi_{U_1}^{(2,2)}$ are shown as the crosses, squares
and asterisks, respectively.}
\label{figU1d}
\end{figure}

In addition, there is a further improvement in using an ancilla dimension
of 3 rather than an ancilla dimension of 2. There are further increases in the
capacity as the ancilla dimension is increased to 4 and 5, as shown in
Fig.\ \ref{figU1d}. These results indicate that the true asymptotic capacity
$\Delta\chi_{U_1}$ is not actually achieved for any particular ancilla dimension.
However, each increase in the capacity with the ancilla dimension is smaller than
the previous, indicating that the results calculated here should give a
good approximation of $\Delta\chi_{U_1}$.

Calculations were also performed for ensembles with 4 members, and ancilla
dimensions up to 4. It was found that, for all ancilla dimensions tested,
there was no increase in the capacity when the number of states in the ancilla was
increased. In addition, it was found that there was no increase in the capacity with
up to 8 states in the ensemble and no ancilla.

\begin{figure}
\centering
\includegraphics[width=0.45\textwidth]{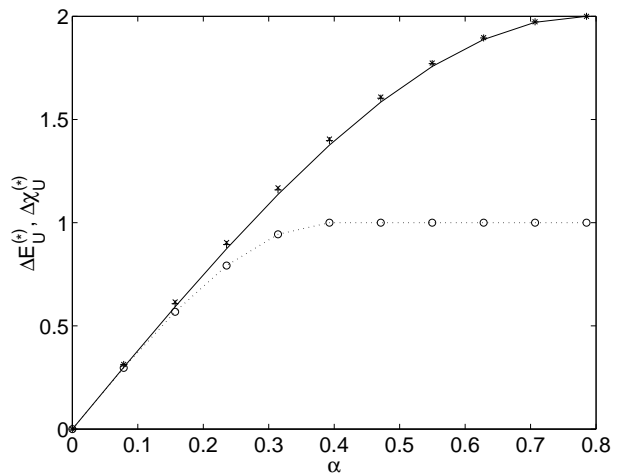}
\caption{The increase in entanglement for arbitrary initial state, $\Delta E_{U_2}^{(*)}$ 
and the increase in Holevo information for arbitrary initial ensemble,
$\Delta\chi_{U_2}^{(*)}$. The values of $\Delta E_{U_2}^{(1)}$
and $\Delta E_{U_2}^{(2)}$ are shown as the dotted and solid lines, respectively, and
the values of $\Delta\chi_{U_2}^{(4,1)}$, $\Delta\chi_{U_2}^{(4,2)}$ and
$\Delta\chi_{U_2}^{(4,3)}$ are shown as the circles, plusses and crosses, respectively.}
\label{figU2b}
\end{figure}

The results for $\Delta\chi_{U_2}^{(*)}$ and $\Delta E_{U_2}^{(*)}$ are shown in
Fig.\ \ref{figU2b}. In each case shown, ensembles with four states were used. In the
case without ancilla, it was found that $\Delta\chi_{U_2}^{(4,1)}$ and $\Delta E_{U_2}^{(1)}$ were equal.
When an ancilla is included, there is a significant increase in both $\Delta\chi_{U_2}^{(*)}$
and $\Delta E_{U_2}^{(*)}$. In particular, these have a maximum of 2, rather than 1 as in the
case without ancilla.

Note also that the value of $\Delta E_{U_2}^{(*)}$ is increased when an ancilla is added for
each of the values of $\alpha$ except the trivial points at $\alpha=0$ and $\pi/4$.
In contrast, for the data shown in Ref.\ \cite{leifer} there was no visible increase
in $\Delta E_{U_2}^{(*)}$ when the ancilla was included for another three data points (at
$\alpha=\pi/40$, $2\pi/40$ and $3\pi/40$). We suspect that this is because the optimization
found the local maximum corresponding to the solution with no ancilla, rather than
the global maximum.

\begin{figure}
\centering
\includegraphics[width=0.45\textwidth]{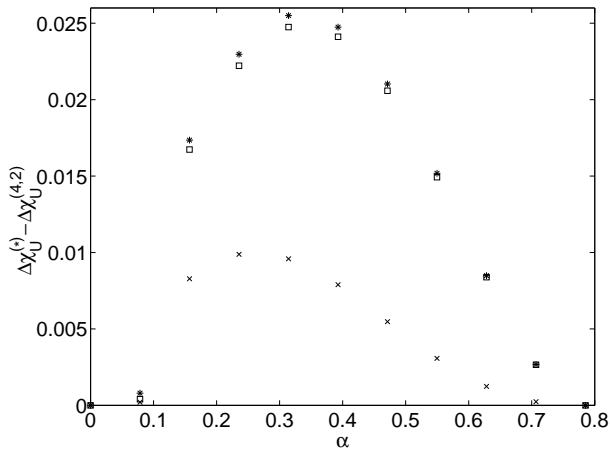}
\caption{The increase in Holevo information for arbitrary initial ensemble,
$\Delta\chi_{U_2}^{(*)}$, relative to that for the case where the ensemble has four members
and the ancilla is of dimension 2, $\Delta\chi_U^{(4,2)}$. The values of
$\Delta\chi_{U_2}^{(4,3)}-\Delta\chi_{U_2}^{(4,2)}$, $\Delta\chi_{U_2}^{(4,4)}-\Delta\chi_{U_2}^{(4,2)}$
and $\Delta\chi_{U_2}^{(4,5)}-\Delta\chi_{U_2}^{(4,2)}$ are shown as the crosses, squares
and asterisks, respectively.}
\label{figU2c}
\end{figure}

We found that using ancilla dimensions above 2 up to an ancilla dimension of 5
did not increase $\Delta E_{U_2}^{(*)}$, in agreement with Ref.\ \cite{leifer}.
When the ancilla was included, $\Delta\chi_{U_2}^{(*)}$ was slightly greater than
$\Delta E_{U_2}^{(*)}$, just as in the case of the operation $U_1$. In addition, it was found
that $\Delta\chi_{U_2}^{(*)}$ was further increased as the ancilla dimension was increased
beyond 2 (see Fig.\ \ref{figU2c}). In this case the difference is somewhat greater,
being around $0.02$ rather than $10^{-4}$, but the values still appear to be
converging for large ancilla dimension. It is expected that the results shown are a
good approximation of the true value of $\Delta\chi_{U_2}$.

\begin{figure}
\centering
\includegraphics[width=0.45\textwidth]{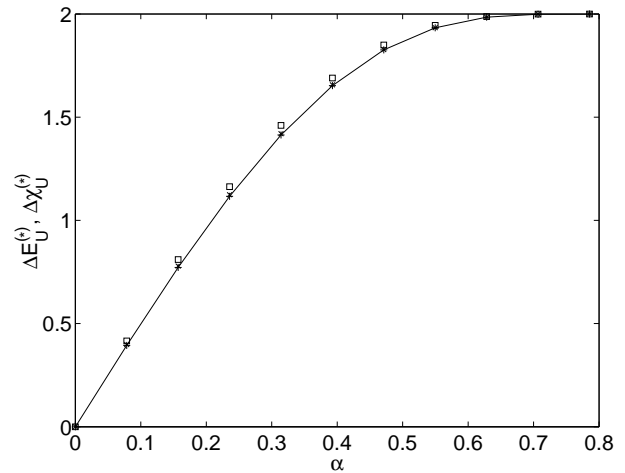}
\caption{The increase in entanglement for arbitrary initial state, $\Delta E_{U_3}$,
and the increase in Holevo information for arbitrary initial ensemble,
$\Delta\chi_{U_3}$. The values of $\Delta E_{U_3}$
are shown as the solid line, and the values of $\Delta\chi_{U_3}^{(4,2)}$,
$\Delta\chi_{U_3}^{(4,3)}$ and $\Delta\chi_{U_3}^{(4,4)}$ are shown as the plusses, crosses
and squares, respectively.}
\label{figU3b}
\end{figure}

The results for the third operation, $U_3$, are shown in Fig.\ \ref{figU3b}. All results
here are for ensembles with four states. In this case it was found that, if the ancilla
had dimension 2, the values of $\Delta\chi_{U_3}^{(4,2)}$ and $\Delta E_{U_3}^{(2)}$ were identical.
In other respects the results were similar to those for the operation $U_2$.
The value of $\Delta E_{U_3}^{(*)}$ was not increased by increasing the ancilla dimension above
2, as for the operations $U_1$ and $U_2$. The value of $\Delta\chi_{U_3}^{(*)}$ was increased
for larger ancilla dimensions, so for these larger ancilla dimensions $\Delta\chi_{U_3}^{(*)}$
was not equal to $\Delta E_{U_3}^{(*)}$.

\begin{figure}
\centering
\includegraphics[width=0.45\textwidth]{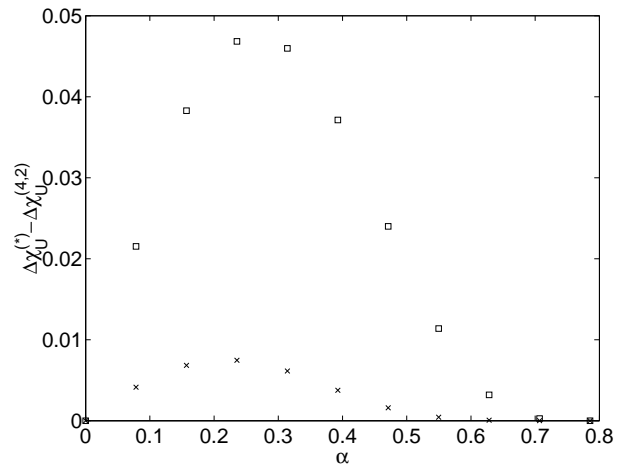}
\caption{The increase in Holevo information for arbitrary initial ensemble,
$\Delta\chi_{U_3}^{(*)}$, relative to that for the case where the ensemble has four members
and the ancilla is of dimension 2, $\Delta\chi_{U_3}^{(4,2)}$. The values of
$\Delta\chi_{U_3}^{(4,3)}-\Delta\chi_{U_3}^{(4,2)}$ and $\Delta\chi_{U_3}^{(4,4)}-\Delta\chi_{U_3}^{(4,2)}$
are shown as the crosses and squares, respectively.}
\label{figU3c}
\end{figure}

The difference between $\Delta\chi_{U_3}^{(*)}$ for higher ancilla dimensions and
$\Delta\chi_{U_3}^{(4,2)}$ is shown in Fig.\ \ref{figU3c}. In this case the difference between
$\Delta\chi_{U_3}^{(4,4)}$ and $\Delta\chi_{U_3}^{(4,2)}$ is almost $0.05$, which is larger than
for both $U_1$ and $U_2$. Nevertheless, the maximum difference between
$\Delta\chi_{U_3}^{(4,4)}$ and $\Delta E_{U_3}^{(2)}$ was still comparable with the results for
$U_1$ and $U_2$.

To summarize, our results strongly indicate that $\Delta\chi_U$ is strictly greater than
$\Delta E_U$ for most two-qubit operations, rather than simply greater than or equal to,
we stated in Refs.\ \cite{berry1,berry2}. In addition, our results show that the maximum change in
Holevo information is not obtained for ancilla dimensions of two, as appears to be the
case for the entanglement. The indications are that there is no ancilla dimension that is
sufficiently large that $\Delta\chi_U$ is achieved, so $\Delta\chi_U$ would only be obtained
asymptotically in the limit of large ancilla dimension.

\section{Conclusions}
\label{concl}
Our results demonstrate that, for two-qubit operations, there are close relationships
between the capacities based on the Holevo information and the capacities based on the
entanglement. In most cases, the largest values calculated for $\chi_U$ (with the
largest ancilla dimension and ensemble size) were equal to $E_U$. In some cases values
of $\chi_U$ were calculated that are slightly above $E_U$, but the difference was very
small. In addition, the maximum values calculated for $\Delta\chi_U$ were very close
to those for $\Delta E_U$. These results indicate that the communication capacities
$\chi_U$ and $\Delta\chi_U$, which are very difficult to calculate, may be estimated 
from the entanglement capacities $E_U$ and $\Delta E_U$, respectively, which may be
calculated far more rapidly.

We have also found that, in all cases tested (except the trivial cases where $\alpha=0$
or $\pi/4$), $\Delta\chi_U$ is greater than $\Delta E_U$, rather than just greater than
or equal to, as stated in Refs.\ \cite{berry1,berry2}. As $\Delta\chi_U$ is equal to
the asymptotic communication capacity in a single direction \cite{bhls}, and the asymptotic
bidirectional communication capacity is no larger than $2\Delta E_U$ for two qubit
operations \cite{berry2}, our results show that there is an inevitable trade-off
involved in performing bidirectional communication. That is, it is not possible to
perform an average communication of $\Delta\chi_U$ in both directions at the same time.

Our results also demonstrate that larger capacities $\chi_U^{(*)}$ and $\Delta\chi_U^{(*)}$
are obtained when the dimension of the ancillas is increased above 2, in contrast to
the case of the entanglement capacities, for which an ancilla dimension of 2 is sufficient.
This has been shown analytically for $E_U$ in Ref.\ \cite{nielsen}, and found numerically
for $\Delta E_U$ in Ref.\ \cite{leifer} (our numerical results also support this).

It must be emphasized that there are limitations on the conclusions that can be drawn from
the results due to their numerical nature. The capacities found here were found by
numerical maximization, which has the drawback that it is possible that, in some cases,
a local maximum may have been found, rather than the global maximum. However, the
calculations presented here were performed very carefully, and in some cases repeated many
times with different random numbers, in order to avoid local maxima. Therefore we are
reasonably confident that the results presented here are very close to the global maxima.

A more serious issue is that it is not possible to perform calculations for arbitrarily
large ancilla dimensions. Although it has been proven that an ancilla
dimension of 2 is sufficient for $E_U$, this has not been proven for $\Delta E_U$. Here
we have found that $\Delta E_U^{(*)}$ is unchanged as the ancilla dimension is raised from
2 to 5. This result strongly indicates that the values obtained for $\Delta E_U^{(*)}$
are the correct asymptotic values.
Nevertheless, it is, in principle, possible that larger values of $\Delta E_U$ may be
achieved for larger ancilla dimensions than those tested. In order to show conclusively that
there is a strict inequality between $\Delta\chi_U$ and $\Delta E_U$, it would be necessary
to prove that the asymptotic entangling capacity is achieved for an ancilla dimension of 2.

In the cases of $\chi_U$ and $\Delta\chi_U$, it is more difficult to estimate the capacity
because the capacity increases with ancilla dimension. For $\Delta\chi_U$ the capacity only
increases by a small amount as the ancilla dimension is increased from 2 to 5, indicating
that the results are a good approximation of the capacity for arbitrary ancilla dimension.
For $\chi_{U_2}$ and $\chi_{U_3}$, there are significant increases in the capacity with the
ancilla dimension up to an ancilla dimension of 4, but there is no increase as the ancilla
dimension is increased to 5. This result indicates that an ancilla dimension of 4 is sufficient,
though further results for higher ancilla dimensions would be required to be confident of this.

In addition there is the problem that it is not possible to perform calculations for
arbitrarily large ensemble sizes. For the operation $U_1$ it was found that the capacity
was not increased as the ensemble size was increased above 2, so it is likely that an
ensemble size of 2 is sufficient for this operation. For the operations $U_2$ and $U_3$ it
was found that the capacities $\chi_U^{(*)}$ and $\Delta\chi_U^{(*)}$ increased with the
ensemble size up to an ensemble size of 4. It is possible that the capacities may be larger
for larger ensemble sizes; this is a topic for future research.

\appendix*
\section{Numerical Methods}
The numerical techniques used to search for the maxima are similar to simulated
annealing \cite{simulated}. For the case of the entanglement capacity $E_U^{(*)}$, a
vector of complex numbers, $(\psi_i)\in\mathbb{C}^{\otimes (\dim{\cal H}_A+
\dim{\cal H}_B)}$, represents the initial tensor product state. The first
$\dim{\cal H}_A$ numbers are the coefficients representing Alice's local state, $\ket{\phi}_A$,
and the last $\dim{\cal H}_B$ numbers are the coefficients representing Bob's local state, $\ket{\chi}_B$.
The total state is simply $\ket{\Psi}=\ket{\phi}_A\otimes\ket{\chi}_B$.
The initial values of $\psi_i$ were selected using a Gaussian distribution
followed by normalization. Throughout this section, all Gaussian distributions
for complex variables are real Gaussians multiplied by random phases (with
a uniform distribution).

At each step another vector of complex random numbers, $(\Delta\psi_i)$, was selected via
a Gaussian distribution.
If the state represented by $(\psi_i+\Delta\psi_i)/{\cal N}$ (where ${\cal N}$ is a
normalization constant) gave a larger final entanglement after application of $U$, then
the coefficients $(\psi_i)$ were replaced with $(\psi_i+\Delta\psi_i)/{\cal N}$.
This technique is equivalent to applying simulated annealing with a temperature of zero,
because in no case were coefficients chosen that gave a smaller final entanglement.
It was found that this technique provided very rapid convergence, and using non-zero
temperatures did not provide better convergence.

Initially the standard deviation in the Gaussian distribution used for the increments, $\sigma$,
was chosen to be 1. The value of $\sigma$ was halved each time there were 1000
consecutive increments tested with none providing a larger final entanglement.
This process was terminated when $\sigma$ fell below $10^{-9}$.
At this stage the progressive changes in the final entanglement were on the order of
$10^{-15}$ or less.

In the case of the capacity $\chi_U^{(*)}$, a vector of complex random numbers, $(\psi_i)
\in\mathbb{C}^{\otimes \dim({\cal H}_A\otimes{\cal H}_B)}$, was used to represent one
initial state in the ensemble, $\ket{\psi}_{AB}$. Initial values of $\psi_i$ were
selected using a Gaussian distribution then normalizing. The remainder of the initial
states in the ensemble were obtained by local unitary operations, $V^{(k)}$. These unitary
operations were represented by matrices of complex numbers, $V_{ij}^{(k)}$. The initial values of
$V_{ij}^{(k)}$ were selected by generating complex random numbers with a Gaussian distribution,
then using Gram-Schmidt orthormalization on the row vectors. The probabilities were 
represented by a vector of real numbers $(p_k)$. Initial values were selected using a uniform
distribution then normalizing.

At each step, complex random numbers for the increments $\Delta\psi_i$ and
$\Delta V_{ij}^{(k)}$ were chosen using a Gaussian distribution with standard deviation
$\sigma$, and real random numbers for the increments $\Delta p_k$ were chosen using a
uniform distribution from $-\sigma/2$ to $\sigma/2$. These increments were used to create
a new ensemble with state $(\psi_i+\Delta\psi_i)/{\cal N}$ and probabilities $p_k(1+\Delta p_k)/
(1+\sum_k p_k \Delta p_k)$. The new operations for the new ensemble were obtained by adding
the increments $\Delta V_{ij}^{(k)}$ to $V_{ij}^{(k)}$ then applying Gram-Schmidt
orthonormalization to the row vectors.

In this case the temperature of the simulated annealing was not taken to be zero. The
new ensemble was selected if the new value for the Holveo information after application
of $U$, $\chi_{\rm new}$, was larger than the previous value, $\chi_{\rm old}$. If the new
value was lower, the new ensemble was selected if
\begin{equation}
R < {\rm e}^{(\chi_{\rm new}-\chi_{\rm old})/\tau},
\end{equation}
where $R$ is a uniform random number between zero and one, and $\tau$ is a tolerance
equivalent to the temperature.
The value of the tolerance $\tau$ was initially taken to be $10^{-6}$. The value of
the final Holevo information was checked every 10000 iterations; if it
had decreased, then the tolerance was divided by 2, and if it had increased, then the
tolerance was multiplied by $1.1$. This adjustment was found to provide reasonably rapid
convergence to the final value.

There were two alternative schemes used to adjust the standard deviation, $\sigma$, in
the Gaussian distribution used for the increments to the ensemble. The first was similar
to that for the entanglement, except that $\sigma$ was halved if 10000 alternatives were
tested with no increase in the Holevo information. The other scheme was to adjust
$\sigma$ such that approximately 20\% of the alternatives were accepted. Only the
first scheme was used in the initial part of the calculation (for the first $10^6$ or
so increments tested). For later parts of the calculation both alternatives were tried.

The numerical techniques used to calculate $\Delta E_U^{(*)}$ and
$\Delta\chi_U^{(*)}$ were similar to those used for $E_U^{(*)}$ and $\chi_U^{(*)}$, though
there are minor differences. In the case of $\Delta E_U^{(*)}$, the initial state,
$\ket{\psi}_{AB}$ was taken to be a general entangled state, rather than a tensor product
of two local states. This state was represented by a vector of coefficients
$(\psi_i)\in\mathbb{C}^{\otimes\dim({\cal H}_A\otimes{\cal H}_B)}$.

The case of $\Delta\chi_U^{(*)}$ is rather simpler than that for $\chi_U^{(*)}$. Rather
than it being necessary to condider a set of unitary operations and probabilities,
the ensemble was simply represented by a set of complex coefficients $\psi_i^{(k)}$
for all of the states in the ensemble. Rather than separately storing probabilities,
these states were not normalized, and the normalizations of these states were taken
to be the probabilities.

\acknowledgments
This research has been supported by an Australian Department of Education Science and
Training Innovation Access Program Grant to support collaboration in the European Fifth
Framework project QUPRODIS, and by Alberta's informatics Centre of Research Excellence
(iCORE).


\begin{thebibliography}{}
\bibitem{nielsen} M. A. Nielsen, C. M. Dawson, J. L. Dodd, A. Gilchrist,
D. Mortimer, T. J. Osborne, M. J. Bremner, A. W. Harrow, and A. Hines, \pra
{\bf 67}, 052301 (2003).
\bibitem{bhls} C. H. Bennett, A. W. Harrow, D. W. Leung, and J. A. Smolin,
quant-ph/0205057 (2002).
\bibitem{berry1} D. W. Berry and B. C. Sanders, \pra {\bf 67}, 040302(R) (2002).
\bibitem{berry2} D. W. Berry and B. C. Sanders, quant-ph/0207065 (2002).
\bibitem{cirac} J. I. Cirac, W. D\"ur, B. Kraus, and M. Lewenstein, \prl
{\bf 86}, 544 (2001).
\bibitem{zanardi} P. Zanardi, C. Zalka, and L. Faoro, \pra {\bf 62}, 030301(R)
(2000).
\bibitem{durvid} W. D\"ur, G. Vidal, J. I. Cirac, N. Linden, and S. Popescu,
\prl {\bf 87}, 137901 (2001).
\bibitem{kradur} B. Kraus, W. D\"ur, G. Vidal, J. I. Cirac, M. Lewenstein,
N. Linden, and S. Popescu, Z. Naturforsch {\bf 56 a}, 91 (2001).
\bibitem{kraus} B. Kraus and J. I. Cirac, \pra {\bf 63}, 062309 (2001).
\bibitem{leifer} M. S. Leifer, L. Henderson, and N. Linden,
\pra {\bf 67}, 012306 (2002).
\bibitem{childs} A. M. Childs, D. W. Leung, F. Verstraete, and G. Vidal,
Quantum Information and Computation {\bf 3}, 97 (2003).
\bibitem{kraus2} B. Kraus, K. Hammerer, G. Giedke, and J. I. Cirac,
\pra {\bf 67}, 042314 (2003).
\bibitem{eisert} J. Eisert, K. Jacobs, P. Papadopoulos, and M. B. Plenio,
\pra {\bf 62}, 052317 (2000).
\bibitem{collins} D. Collins, N. Linden, and S. Popescu, \pra {\bf 64}, 032302
(2001).
\bibitem{holevo} A. S. Holevo, IEEE Trans. Inf. Theory {\bf 44}, 269 (1998).
\bibitem{shuwes} B. Schumacher and M. D. Westmoreland, \pra {\bf 56},
131 (1997).
\bibitem{makhlin} Y. Makhlin, Quantum Information Processing {\bf 1}, 243
(2002).
\bibitem{hammer} K. Hammerer, G. Vidal, and J. I. Cirac, \pra {\bf 66}, 062321
(2002).
\bibitem{simulated} S. Kirkpatrick, C. D. Gerlatt Jr., and M. P. Vecchi,
Science {\bf 220}, 671 (1983).
\end{thebibliography}
\end{document}